\documentclass[aps,prl,twocolumn]{revtex4}
\usepackage{graphicx}
\usepackage{psfrag}

\newcommand{\myscalebox}[1]{\scalebox{0.40}[0.40]{#1}}
\newcommand{\myscaleboxs}[1]{\scalebox{0.22}[0.22]{#1}}
\newcommand{\myscaleboxC}[1]{\scalebox{0.30}[0.30]{#1}}
   
\begin{document}

\title{Long-Time Effects  in a Simulation Model of Sputter Erosion}
\author{Alexander K. Hartmann, Reiner Kree}
\affiliation{Institut f\"ur Theoretische Physik,University of
  G\"ottingen, Bunsenstr. 9, 37073 G\"ottingen, Germany\\
      E-mail: \texttt{hartmann@theorie.physik.uni-goettingen.de}
        }

\author{Ulrich Geyer, Matthias K\"olbel}
\affiliation{1. Physikalisches Institut, University of
  G\"ottingen, Bunsenstr. 9, 37073 G\"ottingen, Germany}
\date{\today}

\begin{abstract}

 A simple (2+1) dimensional discrete 
  model is introduced to study the evolution of
 solid surface morphologies during ion beam sputtering.
 The model is based on the same assumptions about the erosion
 process as the existing analytic theories.
 Due to its simple structure, simulations of the model 
 can be performed on time scales, where effects beyond
 the linearized theory become important.
 Whereas for short times we observe the formation of ripple structures
 in accordance with the linearized theory,  we find a roughening surface 
 for intermediate times. 
 The long time behavior of the model strongly depends on the surface
 relaxation mechanism.

{\bf Keywords (PACS-codes)}: 
68.35.-p, 
79.20.Rf, 
82.20.Wt 
\end{abstract}
\maketitle

\paragraph*{Introduction}

During the last years, two features of surface morphologies created by ion
beam sputtering attracted particular attention:
ripple structures on sub-micrometer length scales  and self-affine,
rough surfaces\cite{barabasi1995}.

The formation of periodic ripple
structures  has been observed experimentally in amorphous materials
\cite{mayer1994}, metallic crystals \cite{rusponi1998,habenicht1999} 
and semiconductors amorphized by the ion beam \cite{lewis1980,chason1994}. 
Ripples are typically oriented perpendicular to the projection of
the ion beam in the surface plane for  small angles of incidence $\Theta$ (relative to  the surface
normal), whereas for larger angles $\Theta$, the observed ripple pattern is rotated by $90^{\circ}$.
Surfaces eroded by ion bombardment may also exhibit self-affine properties \cite{eklund1991}. 
With increasing ion fluence, a crossover from ripple structures to self-affine, 
rough surfaces has recently been observed experimentally \cite{habenicht1999}.  

Our present understanding of these features is based upon the work of  
Bradley and Harper (BH) \cite{bradley1988}, who found that Sigmund`s
sputtering theory \cite{sigmund1969}
implies a curvature dependence of the sputtering yield.
Based on BH a continuum theory of surface evolution by sputter erosion was formulated as an
anisotropic  Kuramoto-Sivashinsky (KS) equation \cite{KS}
with additive
noise  \cite{cuerno1995b,makeev1997,carter1999}. 
However, there are strong indications from experiment \cite{rusponi1998, rusponi1998b} that surface relaxation processes, 
are also important
during pattern formation. Such processes have  not yet been adequately
included in existing theories. 
Our results will show that the long-time behavior of patterns
depends crucially on details of surface relaxation.

To investigate the analytic theory beyond the linearized regime, 
numerical integrations of the KS equation have been performed \cite{rost1995} \cite{park1999}, which uncovered 
two markedly different long-time regimes, depending on the signs of the non-linear couplings.

Computer simulations may be helpful in clarifying both the role of surface relaxation and of non-linear effects. 
Two types of simulations have been performed up to now. 
Koponen et al  \cite{koponen1996, koponen1997} calculated collision cascades emerging from single ion impact within
the binary collision approximation. They find 
ripples in accordance 
with linearized BH theory \cite{koponen1997}, which appear both with and  
without additional surface relaxation processes.
This indicates the presence of an ion-induced surface diffusion mechanism, 
which has also been predicted from BH \cite{carter1999}.
The simulations did not yet reach time scales where the non-linear
effects of the continuum theories could be analyzed. On the other hand, 
scaling properties of the roughness of ion-irradiated surfaces 
 have also been investigated within this approach \cite{koponen1996}.
In a different approach, Cuerno et al \cite{cuerno1995a} proposed a simple, discrete stochastic 
model with an update rule, 
which incorporates the $\Theta$-dependence of the sputtering yield and 
a simple curvature dependence of the erosion probability {\em ad hoc}.
Within this model, it is possible to study the crossover from ripples to rough surfaces 
during the evolution of an irradiated 1-dimensional system.

In the present letter, we introduce a simple, yet atomistic Monte Carlo
model on a lattice, which includes the same assumptions on energy deposition from ion impact 
as BH  and we study the evolution of surface morphologies beyond the linearized regime for 
two different types of surface relaxation mechanisms.

\paragraph*{Model} 
We model energy deposition by directly implementing the result of
Sigmund's 
sputtering theory \cite{sigmund1969}, upon 
which BH and the existing analytical theories are based. 
The surface is described as a two-dimensional field of
discrete time-dependent 
height variables $h(x,y,t)$ on a square lattice of size $L\times L$. $t$ denotes
the time, measured in terms of ion fluence.
In our results, we also indicate the corresponding amount of eroded
material in terms of eroded monolayers (ML).
Periodic boundary conditions are assumed, i.e.\
$h(x+L,y,t)=h(x,y,t)=h(x,y+L,t)$. Initially the surface is flat,
i.e. $h(x,y,0)=h_0$.
Two processes take place: erosion and
surface relaxation.
For each erosion step an ion is started at a
random position in $[0,L]\times[0,L]$ above the surface and moved
towards the surface with incidence angle $\Theta$ and an angle
$\phi$ with respect to the $x$-axis. After the ion has penetrated a
distance $a$ under the surface it stops and distributes its energy.
A particle at the surface obtains the energy given by  Sigmund`s
theory \cite{sigmund1969} :
  \begin{equation}
    \label{eq:sigmund}
    E(x^{\prime}, y^{\prime},z^{\prime}) =
    \frac{\epsilon}{(2\pi)^{3/2}\sigma\mu^2}
    \exp \left( -\frac{(z^{\prime}+a)^2}{2\sigma^2} 
            -\frac{x^{\prime 2}+y^{\prime 2}}{2\mu^2} \right)
  \end{equation}
where $(x^{\prime},y^{\prime}, z^{\prime})$ is the position of a particle
in a local cartesian coordinate system of the ion, where the $z^{\prime}$ axis
coincides with the ion trajectory.
A particle at point ${\bf r}=(x,y,h)$ on the surface is removed,
i.e. the height variable 
decreased by one, with a probability proportional to $E({\bf r})$
(see  Fig. \ref{figModel}).
 
\newcommand{\captionModel}
{The model consists of a square field  of discrete height
  variables $h(x,y)$, corresponding to piles of $h(x,y)$ particles at position
  $(x,y)$. {\em Left:} Each ion impact is modeled by an distribution describing
  the energy deposited by the ion. Atoms on the surface are removed
  with a probability proportional to the energy. {\em Right:} Surface
  diffusion, by decreasing height differences the energy is decreased.
}

 \begin{figure}[htb]
 \begin{center}
 \myscaleboxC{\includegraphics{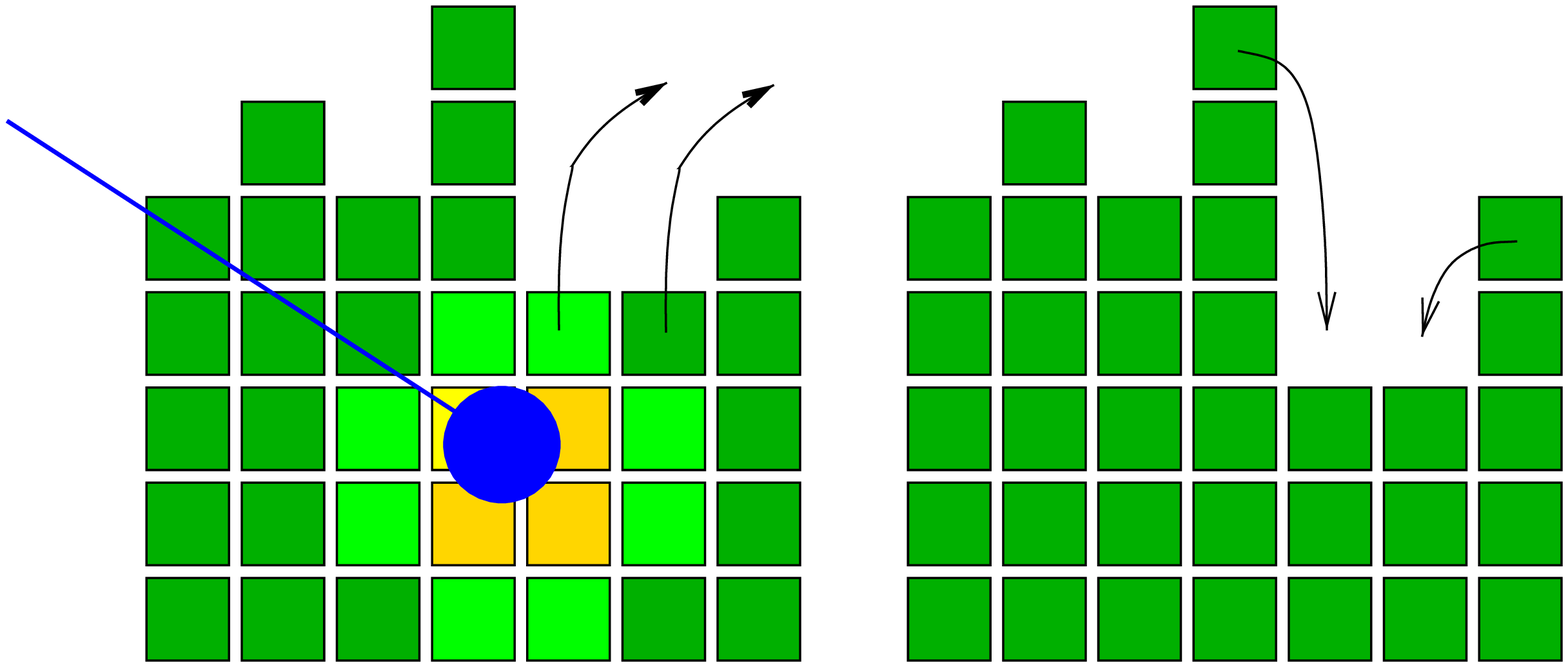}}
 \end{center}
 \caption{\captionModel}
 \label{figModel}
 \end{figure}

For surface relaxation, we use either activated surface diffusion
(ASD) or Wolf-Villain (WV) 
type irreversible relaxation as simplified models of high-temperature
and low-temperature relaxation mechanisms.
ASD is implemented as a nearest neighbor MC hopping process 
with an energy $E=(J/2)\sum_{<i,j>}(h(i)-h(j))^2$ as in 
Ref. \onlinecite{siegert1994} (see also 
right part of Fig. \ref{figModel}) and 
WV relaxation is implemented
according to Ref. \onlinecite{wolf}.
For ASD a MC step is performed every $0.001 L^2$ erosion steps
with temperature $k_bT/J=0.2$. The irreversible WV process is completed for all surface 
atoms after every ion impact. System sizes up to $L = 256$
are considered. To avoid anisotropies, which may be induced by
the square lattice, we have chosen the azimuthal angle
$\phi=22^{\circ}$. The angle $\Theta$ was varied between $0^{\circ}$ and $80^{\circ}$.

We use parameters $a=5.4$, $\sigma=3.3$ and
$\mu=1.7$ and  $\epsilon=(2\pi)^{3/2}\sigma\mu^2$ (which makes the prefactor
in Eq. (\ref{eq:sigmund}) equal to one).   
One should keep in mind that our choice of $\epsilon$ leads to rather high sputtering yields $Y (Y\approx 7.0)$ 
compared to experiments like \cite{habenicht1999}, where $Y=0.3\ldots 0.5$.
According to BH, the ripple wavelength $\lambda$ scales like 
$\lambda \propto Y^{-1/2}$ so that in our simulations we expect patterns with correspondingly 
smaller length scales.  

\paragraph*{Results}
The model perfectly reproduces the $\Theta$ dependence of the sputtering yield of BH \cite{unpublished}  and 
thus should not be used for too large $\Theta$.   
Wavelength ratios obtained from simulations ( $\lambda(45^{\circ})/\lambda(30^{\circ})=0.89(10)$, 
$\lambda(60^{\circ})/\lambda(30^{\circ})=0.84(10)$,  
$\lambda(70^{\circ})/\lambda(30^{\circ})=0.89(10)$ agree reasonably well with linear BH theory 
and with experiment \cite{habenicht1999}.  

Next, we present results on the evolution of the morphology of a 
surface relaxed by ASD to discuss the crossover between different
time regimes. Afterwards we will show that the morphologies reached with WV
relaxation are completely different.
 
For ASD, real space pictures of the height profiles of a $256\times 256$
surface irradiated  with ions under 
$\Theta=50^{\circ}$ are shown in
Fig. \ref{figImages} for increasing ion fluences.  Fig. \ref{figSK} shows 
structure
factors $S({\bf k})=|\hat{h}({\bf k})|^2$ for wavevectors $k_{\parallel}$ 
parallel and $k_{\perp}$
perpendicular to the projection of the direction of the ion beam onto
the flat surface. Here $\hat{h}({\bf k})$ is the Fourier transform of
height fluctuations $(h(x,y)$ $-\bar{h})$ ($\bar{h}=\sum_{x,y}h(x,y)/L^2$ being the
average height).
First signs of a  structure  appear after 
1 ML of eroded material. 
For short times, after three MLs
have been sputtered ($t=2.7\times10^4$ ions), 
a pattern with a typical
length scale of 7 lattice spacings can be seen in the real
space pictures.
After 30 MLs have been eroded ($t=2.5\times 10^5$)
ripples oriented perpendicular to the direction of the ion beam
are visible and maxima in the corresponding structure factor
appear.
For intermediate times ($t=2.6\times 10^6$ corresponding to 300 MLs),
the ripples start to disappear again. After long times ($t=9.3\times 10^7$
or $10^4$ MLs) the morphology approaches 
a rough surface with anisotropic statistics of height
fluctuations. In the structure factor, the
 peak indicating the ripples has strongly decreased and $S(k_{\parallel})$ 
approaches a scaling behavior, $S(k_{\parallel})\sim k_{\parallel}^{-2.83}$.
A similar crossover from ripples to a rough surface is also seen in
experimentally obtained structure factors \cite{habenicht1999}. Please note
that we also have tested that when switching of the sputtering mechanism, the
structure factor obtained from  pure surface diffusion follows at
long times the usual $S(k)\sim k^{-2}$ behavior\cite{barabasi1995}.

\newcommand{\captionImages}
{Surface of a $256\times 256$ system after $t$ ions have hit the system.
  $\Theta=50^{\circ}$. Upper
  row left: $t=2.7\times10^4$ (corresponding to 3 MLs of particles
  been eroded), upper row right: $t=2.5\times 10^5$ (30 MLs), 
lower row left: $t=2.6\times 10^6$ (300 MLs),
lower row right: $t=9.3\times 10^7$ ($10^4$ MLs). 
The bars indicate the direction of the ion beam.}

\begin{figure}[htb]
\begin{center}
\myscaleboxs{\includegraphics{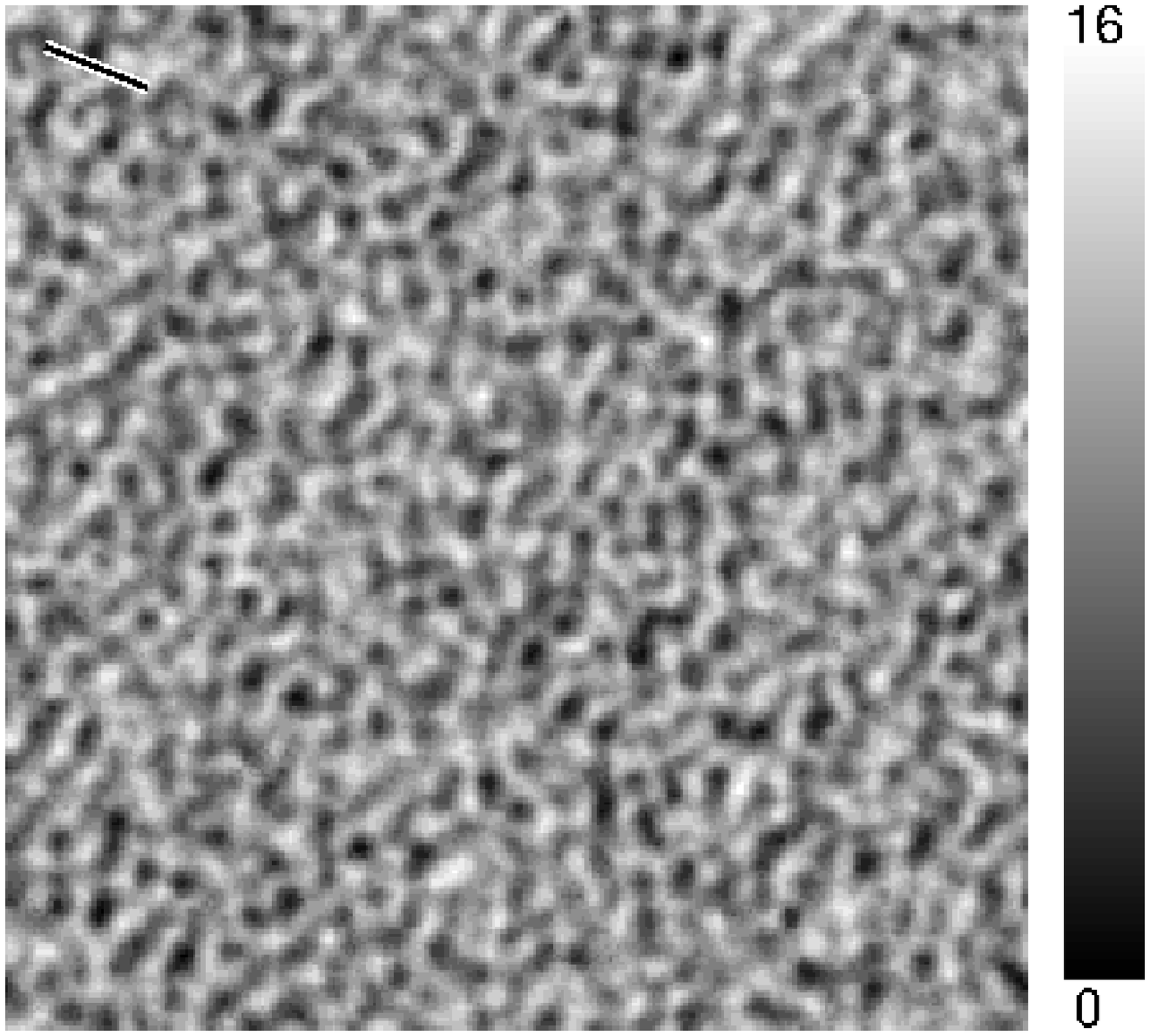}}
\myscaleboxs{\includegraphics{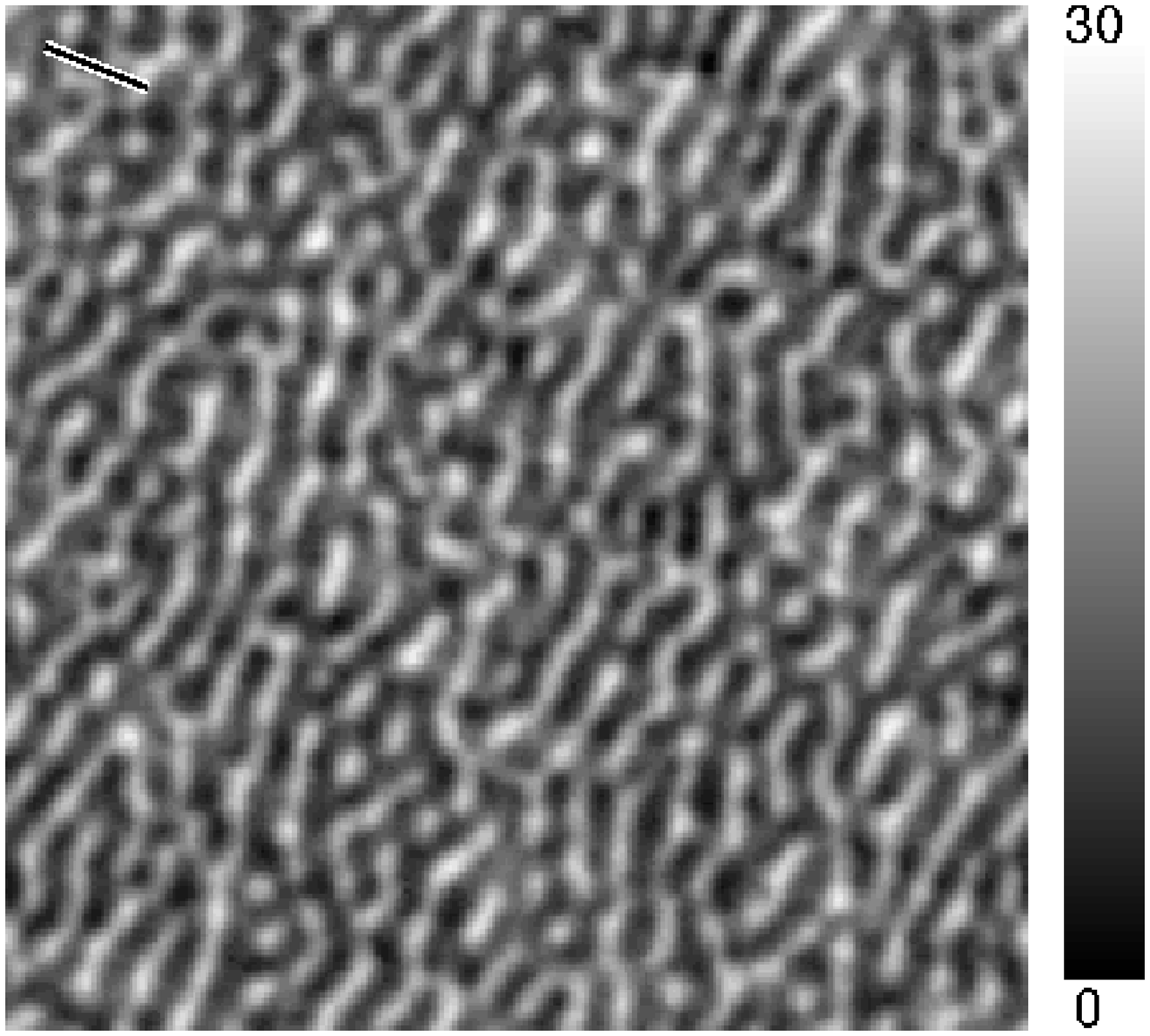}}
\end{center}
\begin{center}
\myscaleboxs{\includegraphics{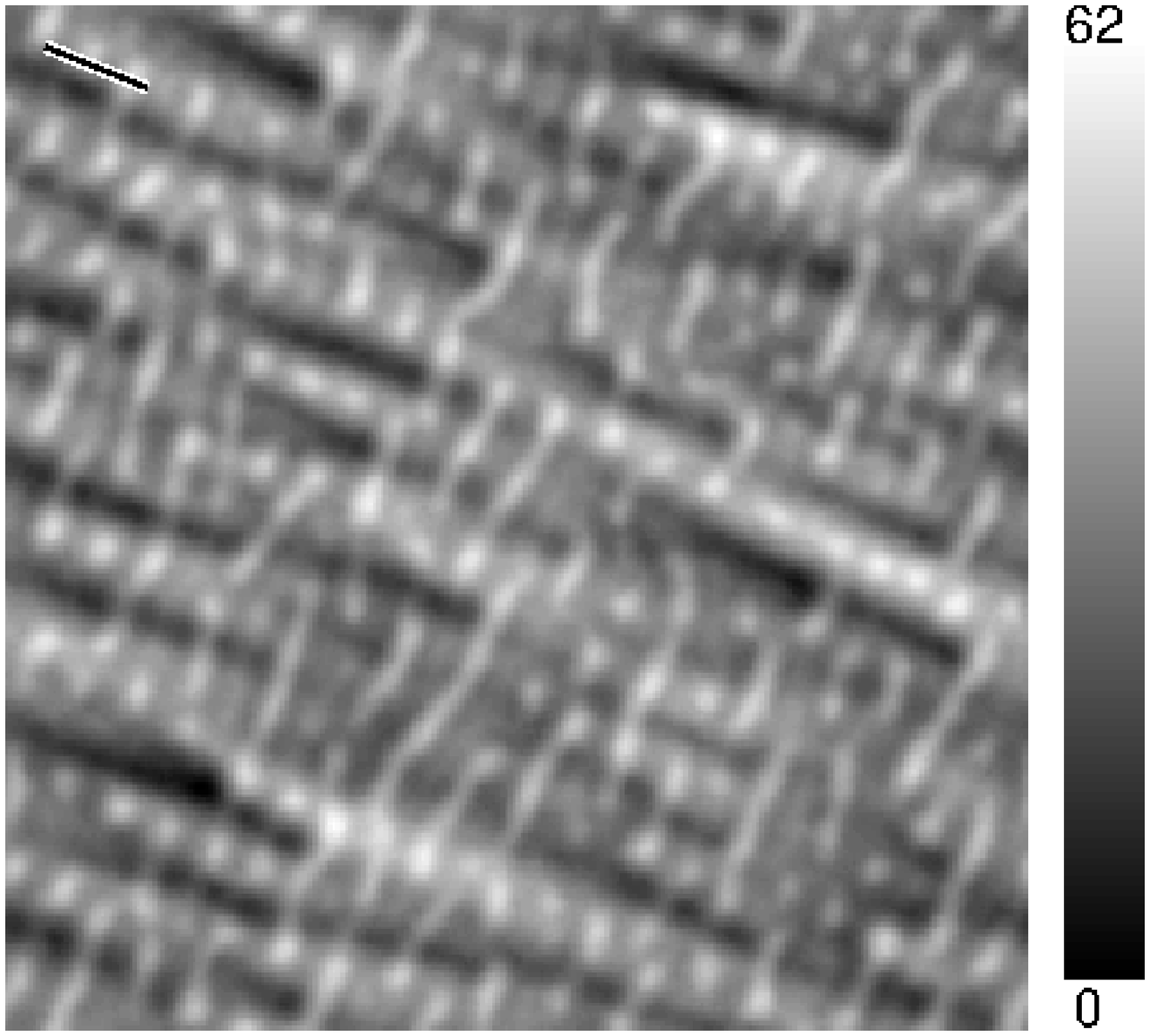}}
\myscaleboxs{\includegraphics{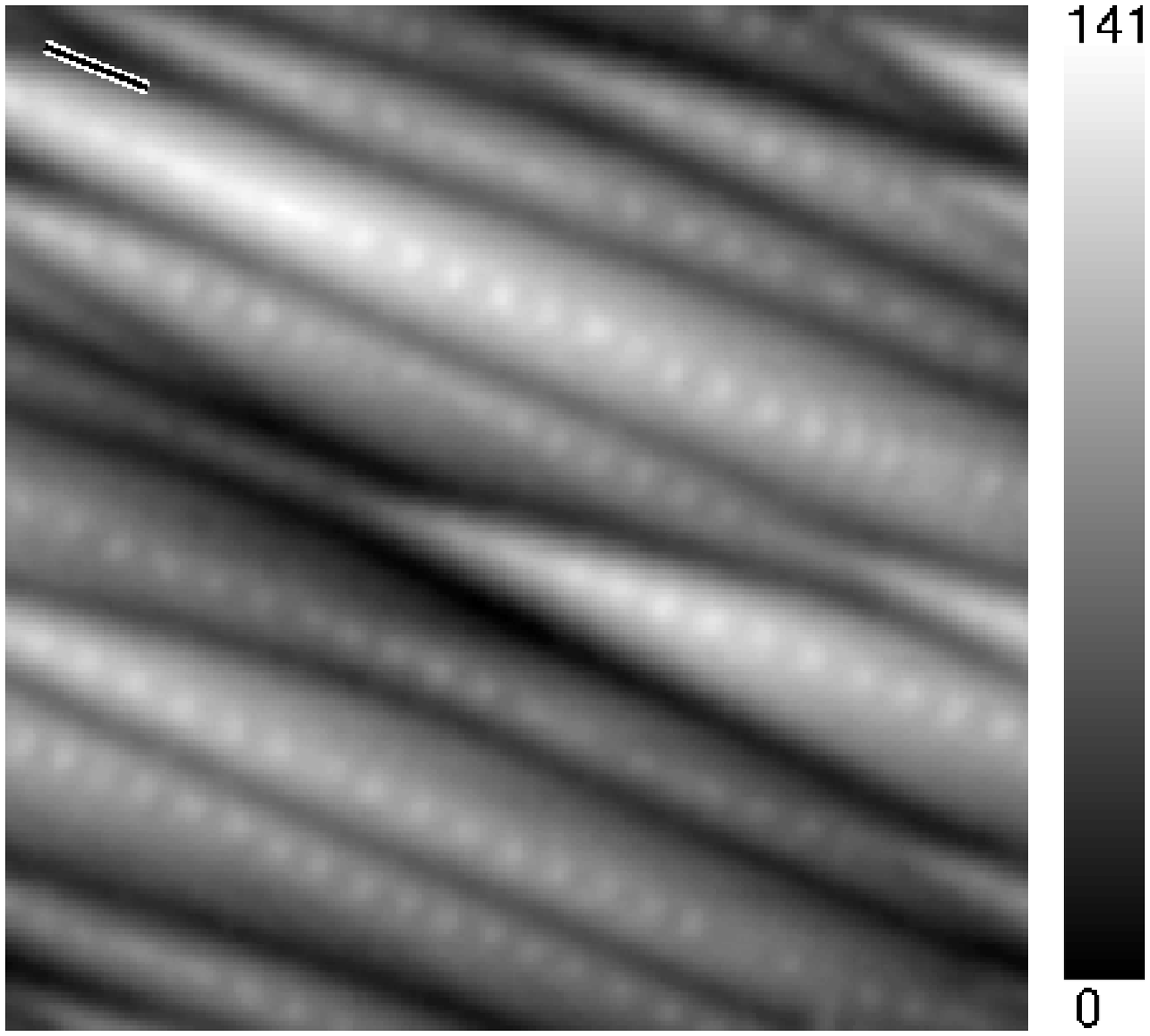}}
\end{center}
\caption{\captionImages}
\label{figImages}
\end{figure}

\newcommand{\captionSK}
{Structure factor $S(k_{\parallel})$ (upper panel) and
$S(k_{\perp}) $ (lower panel) for different ion fluences and  
for $\Theta=50^{\circ}$. The solid line in the upper panel represents
a $k^{-2.83}$ behavior. The data for the largest ion fluence is an
average over 10 independent runs, while the other data points are from
300 runs.}

\begin{figure}[htb]
\begin{center}
\myscalebox{\includegraphics{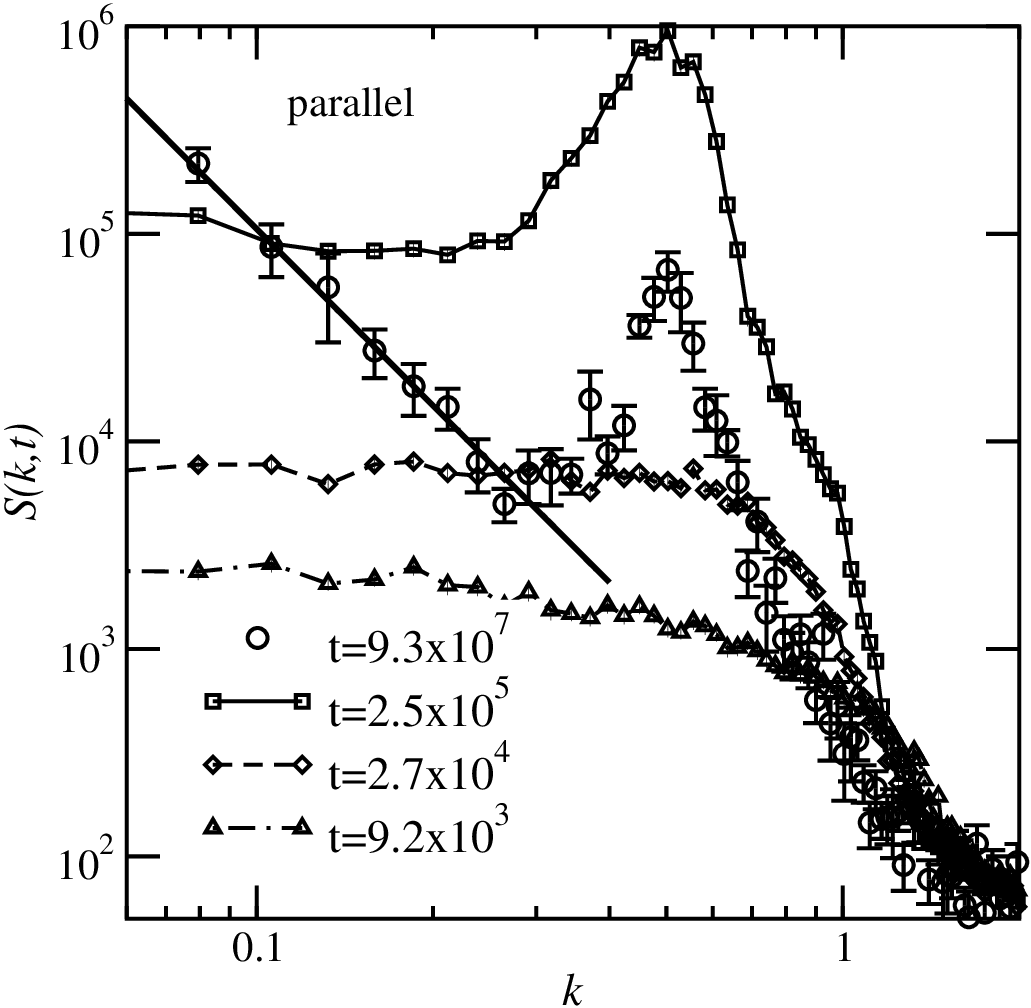}}
\myscalebox{\includegraphics{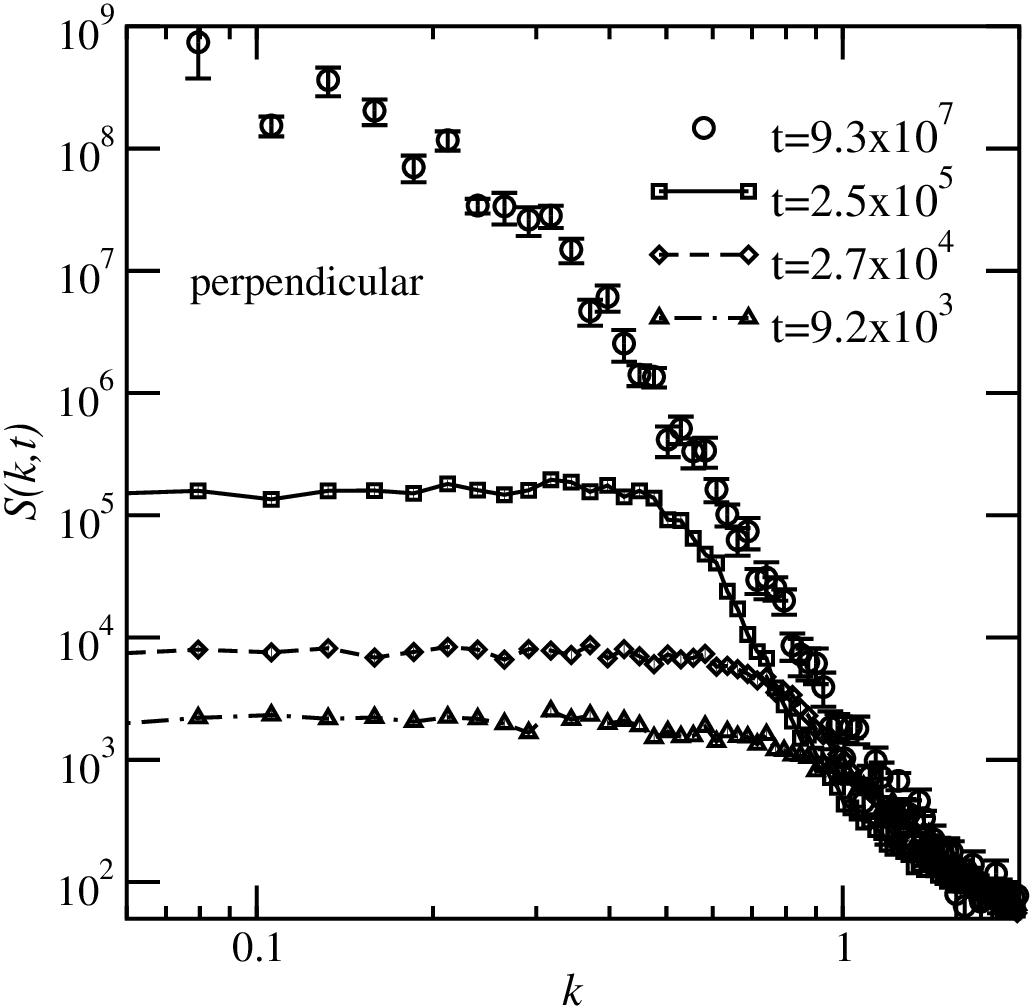}}
\end{center}
\caption{\captionSK}
\label{figSK}
\end{figure}

The different time regimes are seen most clearly in the time evolution of the 
roughness $W^2(t)=1/L^2 \sum_{x,y}(h(x,y,t)-\bar{h}(t))^2$ (see Fig. \ref{figWidth}).
For short
times the data exhibit a positive curvature. 
During that period the ripple structure
is formed. The crossover to intermediate times appears rather abruptly
in accordance with numerical solutions of the noisy KS equation
\cite{park1999}.  
A fit with an algebraic growth law $W \sim t^{\beta}$ for
intermediate times reveals $\beta = 0.168(2)$.
For longest times, a strong increase in $W$ is observed.
We assume that the end of the algebraic regime is marked by finite size effects,
which remain to be studied in detail.

\newcommand{\captionWidth}
{Surface roughness $W^2$ as a function of number $t$ of ions 
  for $\Theta=50^{\circ}$.}

\begin{figure}[htb]
\begin{center}
\myscalebox{\includegraphics{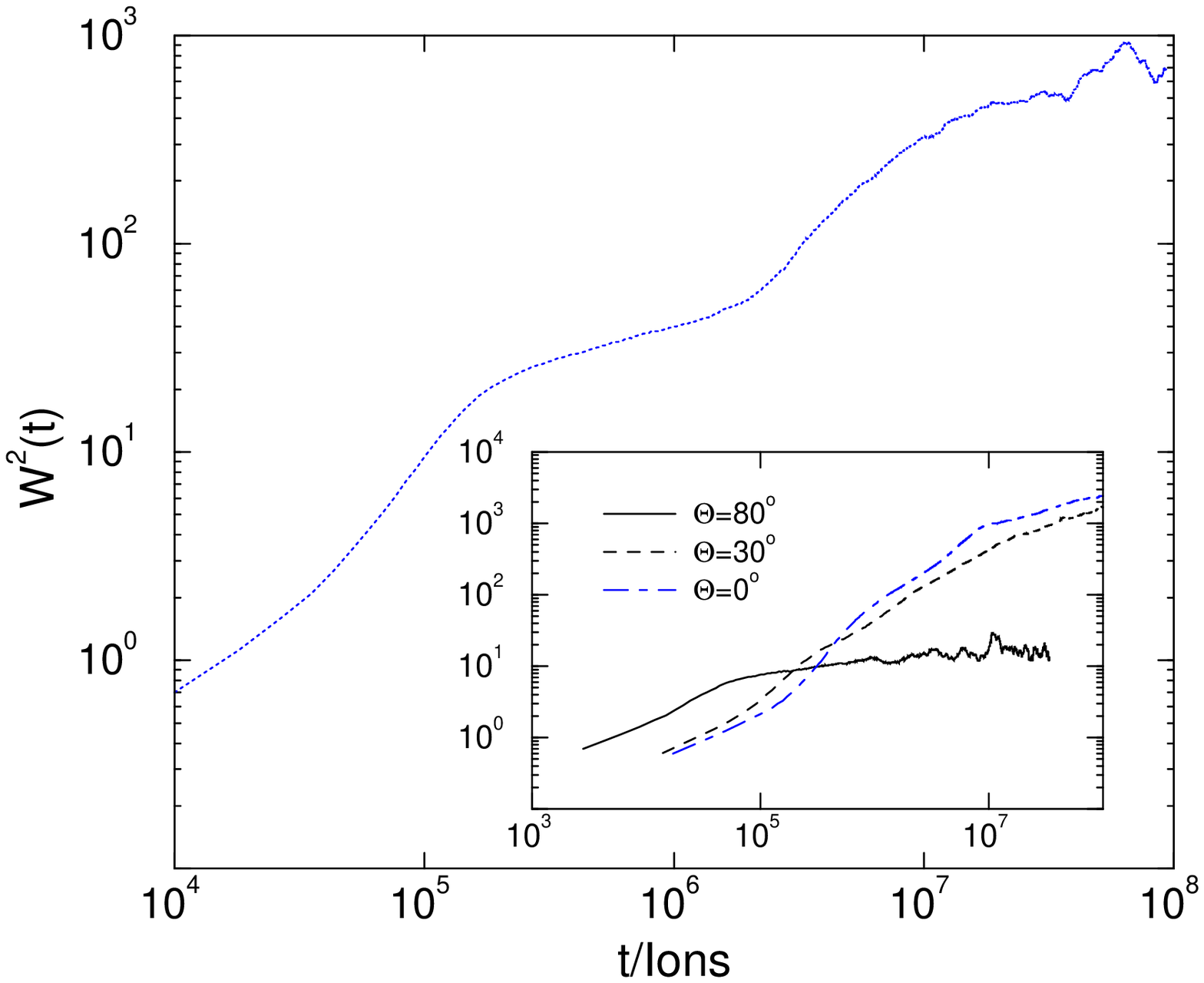}}
\end{center}
\caption{\captionWidth}
\label{figWidth}
\end{figure}

For WV relaxation and for short times, we again find ripples 
which behave in accordance with linear BH theory. The long-time 
morphologies, however, 
are completely different from ASD relaxed surfaces as can be seen in
Fig. \ref{figWolfVillain}. Generally, WV produces stable patterns of pronounced ripples
up to the longest times simulated.
For $\Theta=50^{\circ}$ the ripples are tilted by $\approx 17^{\circ}$ against the
direction perpendicular to incidence, whereas for $\Theta=70^{\circ}$ striplike domains oriented approximately
parallel to incidence appear, which contain ripples with the same tilt angle of $\approx 17^{\circ}$.

\newcommand{\captionWolfVillain}
{Surfaces after long time sputtering with Wolf-Villain type surface diffusion.
left: $\Theta = 50^{\circ}$ right: $\Theta = 70^{\circ}$}

\begin{figure}[htb]
\begin{center}
\myscaleboxs{\includegraphics{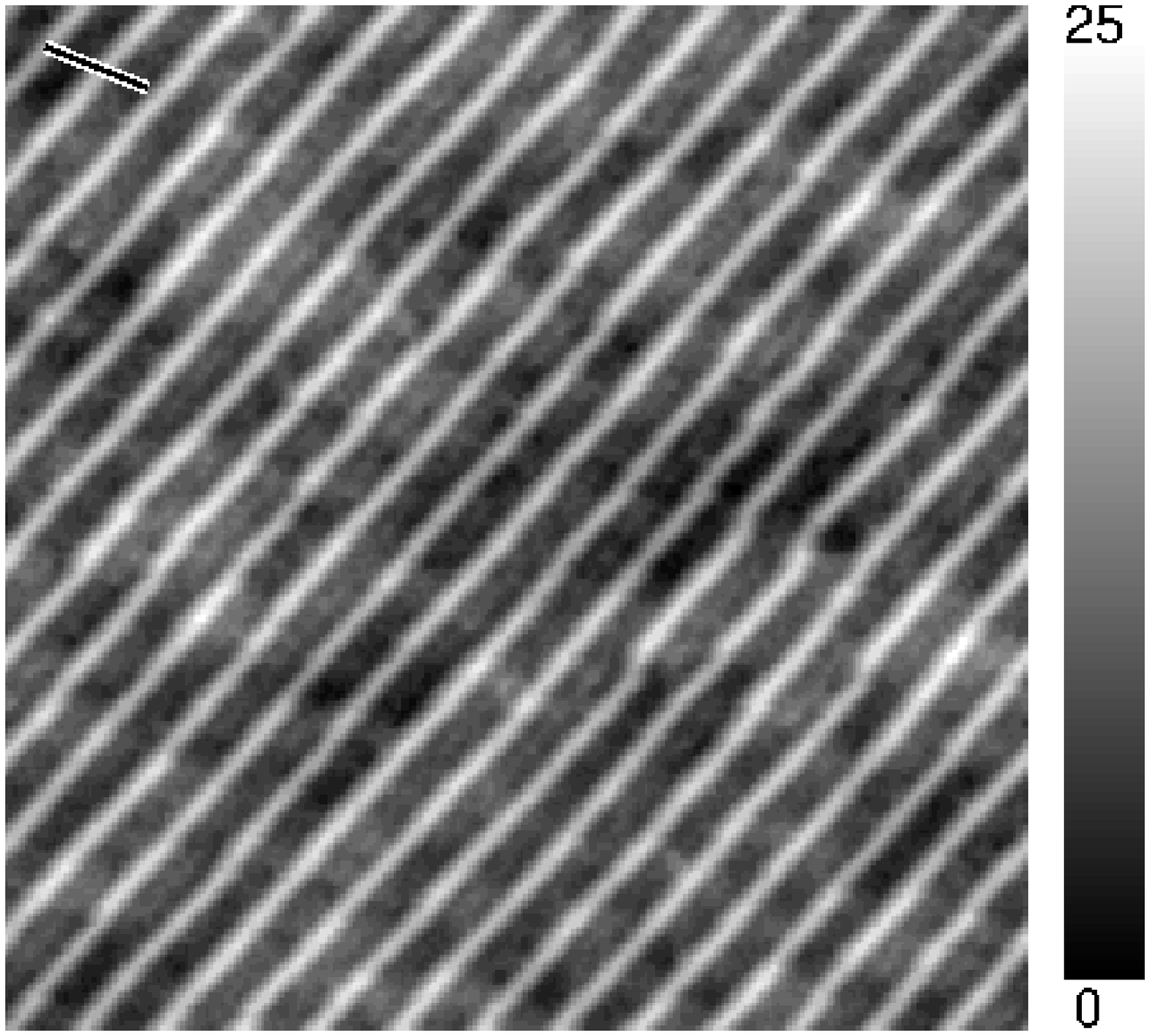}}
\myscaleboxs{\includegraphics{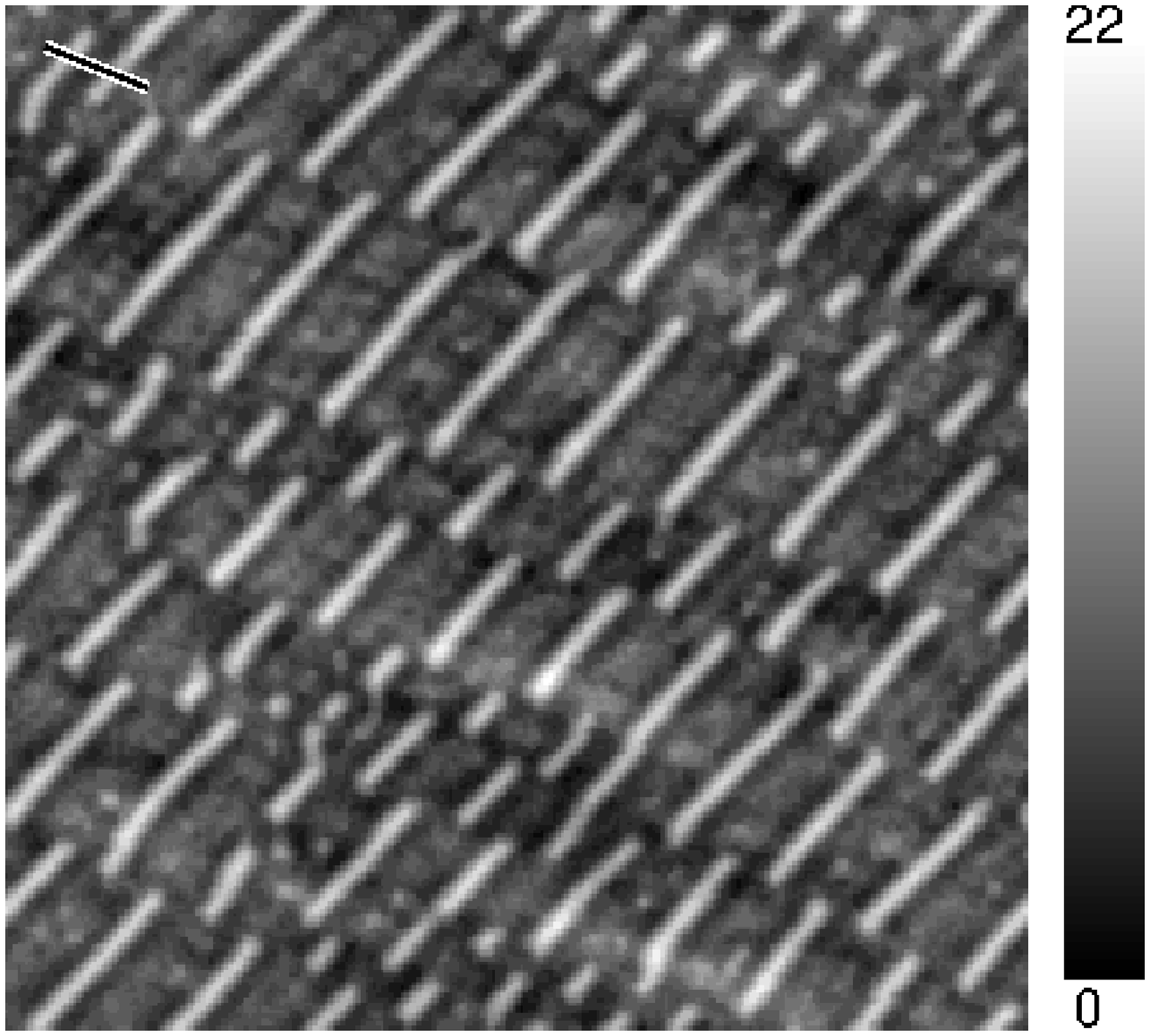}}
\end{center}
\caption{\captionWolfVillain}
\label{figWolfVillain}
\end{figure}



In conclusion, we have 
presented a simple discrete model for (2+1)-dimensional
surfaces subjected to ion-beam sputtering. The model is based on assumptions,
which also underly BH theory and continuum theories.
The simplicity of the model allows us to
simulate rather large systems over time scales, where non-linear effects
become apparent. We have used the model to study distinct time regimes
of surface evolution under the influence of two different surface
relaxation mechanisms (ASD and WV). For short times, we always find
ripple formation in accordance with linear BH theory. 

For longer times, the surface morphologies emerging
from ASD and from WV become qualitatively different. 
For ASD, the ripple wavelength increases with increasing ion fluence,
and  scaling behavior in the structure factor appears. Ripples get
blurred and the surface morphology becomes rough, but stays anisotropic. 
Replacing ASD by WV  surface relaxation, we find 
pronounced ripples in the long-time regime. Please note that in the
earlier detailed simulations of the sputtering process
\cite{koponen1997}, basically the same result was obtained even when
no surface diffusion was present. It was assumed that this is due
to the smoothing mechanism introduced by the detailed modeling
of the sputtering itself. In our simple model, we need an additional
surface relaxation process to observe ripples, but we have found for
short times no significant differences between the WV and ASD cases. Hence, the
non-linear long time regime allows a much better comparison of
different surface diffusion mechanisms.

WV relaxation and ASD may be considered as the simplest type of relaxation
mechanisms for low temperatures and for high temperatures, respectively. 
Please note that the aim of this work is not the evaluate the
relaxation mechanisms itself, but to introduce a simple model which
allows to study long-time effects and their dependence of the surface
diffusion mechanism.
Other, more specific surface relaxation mechanisms
\cite{das-sarma1991,wilby1992,siegert1992,smilauer1993,ramana-murty1999,%
malarz1999,ghaisas2001,pundyindu2001} 
involving diffusion
barriers  may be studied within our model using appropriate kinetic MC
steps for surface relaxation, especially the model can be easily extended
to describe different crystal structures and include surface anisotropies.

AKH announces financial support from the DFG ({\em Deutsche 
Forschungsgemeinschaft}) under grant Zi209/6-1.


\begin{thebibliography}{99}
\bibitem{barabasi1995} A.-L. Barab\'asi and H.E. Stanley, {\it Fractal
concepts in surface growth}, (Cambridge University Press, Cambridge 1995)
\bibitem{mayer1994} T.M. Mayer, E. Chason, and A.J. Howard,
  J. Appl. Phys. {\bf 76}, 1633 (1994)
\bibitem{rusponi1998} S. Rusponi, G. Costantini, C. Boragno, and
  U. Valbusa, Phys. Rev. Lett. {\bf 81}, 4184 (1998)
\bibitem{habenicht1999} S. Habenicht, W. Bolse, K.P. Lieb, K. Reimann,
  and U. Geyer, Phys. Rev. B {\bf 60}, R2200 (1999)
\bibitem{lewis1980} G.W. Lewis, M.J. Nobes, G. Carter, and
  J.L. Whitton, Nucl. Instrum. Methods {\bf 170}, 363 (1980)
\bibitem{chason1994} E. Chason, T.M. Mayer, B.K. Kellerman,
  D.T. McIlroy, and A.J. Howard, Phys. Rev. Lett {\bf 72}, 3040 (1994)
\bibitem{eklund1991} E.A. Eklund et al Phys. Rev. Lett. {\bf 67}, 1759 (1991) 
\bibitem{bradley1988} R.M. Bradley, J.M. Harper,
  J. Vac. Sci. Technol. A {\bf 6}, 2390 (1988)
\bibitem{sigmund1969} P. Sigmund, Phys. Rev. {\bf 184}, 383 (1969)
\bibitem{KS} Y. Kuramoto and T. Tsuzuki, Prog. Theor. Phys. {\bf 55},  356 (1977);
 G.I. Sivashinsky, Acta Astronaut. {\bf 6}, 659 (1979)
\bibitem{cuerno1995b} R. Cuerno and A.-L. Barab\'asi,
  Phys. Rev. Lett. {\bf 74}, 4746 (1995) 
\bibitem{makeev1997}  M. Makeev and A.-L. Barab\'asi, 
Appl. Phys. Lett. {\bf 71}, 2800 (1997)
\bibitem{carter1999} G. Carter, Phys. Rev. B {\bf  59}, 1669 (1999)
\bibitem{rusponi1998b} S. Rusponi, G. Costantini, C. Boragno, and
  U. Valbusa, Phys. Rev. Lett. {\bf 81}, 2735 (1998)
\bibitem{rost1995} M. Rost and J. Krug, Phys. Rev. Lett. {\bf 75} , 3894 (1995)
\bibitem{park1999} S. Park, B. Kahng, H. Jeong and A.-L. Barab\'asi,
    Phys. Rev. Lett. {\bf 83}, 3486 (1999)
\bibitem{koponen1996} I. Koponen, M. Hautala, O.-P. Siev\"anen,
  Phys. Rev. B {\bf 54}, 13502 (1996) 
\bibitem{koponen1997} I. Koponen, M. Hautala, and O.-P. Siev\"anen,
  Phys. Rev. Lett. {\bf 78}, 2612 (1997) 
\bibitem{cuerno1995a} R. Cuerno, H.A. Makse, S. Tomassone,
  S.T. Harrington, and H.E. Stanley, Phys. Rev. Lett. {\bf 75}, 4464 (1995)
\bibitem{siegert1994} M. Siegert and M. Plischke, Phys. Rev. E {\bf
    50}, 917 (1994)
\bibitem{wolf} D.E. Wolf and J. Villain, Europhys. Lett. {\bf 13}, 389 (1990)
\bibitem{unpublished} A. Hartmann, M. K\"olbel, M. Feix, U. Geyer 
and R. Kree, {\em unpublished}
\bibitem{das-sarma1991} S. Das Sarma and P. Tamborenea,
Phys. Rev. Lett. {\bf 66}, 325 (1991)

\bibitem{wilby1992} M.R. Wilby,  D.D. Vvedensky, and  A. Zangwill,
Phys. Rev. B {\bf 46}, 12896 (1992); {\bf 47}, 16068 (E) (1993)

\bibitem{siegert1992} M. Siegert and M. Plischke,
Phys. Rev. Lett. {\bf 68}, 2035 (1992)

\bibitem{smilauer1993} P. Smilauer, M.R. Wilby, and D.D. Vvedensky,
Phys. Rev. B {\bf 47}, 4119 (1993)

\bibitem{ramana-murty1999}     M.V. Ramana Murty and B.H. Cooper,
Phys. Rev. Lett. {\bf 83}, 352 (1999)

\bibitem{malarz1999} K. Malarz and A.Z. Maksymowicz, 
Int. J. Mod. Phys. C {\bf 10}, 659 (1999)

\bibitem{ghaisas2001} S.V. Ghaisas, Phys. Rev. E {\bf 63}, 062601 (2001)

\bibitem{pundyindu2001}
      P. Punyindu, Z. Toroczkai, and S. Das Sarma,
Phys. Rev. B {\bf 64}, 205407 (2001) 

\end{thebibliography}
\end{document}